\documentclass[12pt]{article}
\usepackage{amsmath}
\usepackage{graphicx}

\newcommand{\br}{{\mathbf r}}
\newcommand{\pF}{p_{\mathrm F}}
\newcommand{\EF}{E_{\mathrm F}}

\begin{document}
\begin{center}
{\bfseries Majorana: from atomic and molecular, to nuclear
physics}\\[\baselineskip]
R. Pucci and G. G. N. Angilella\\[\baselineskip]
\begin{em}
Dipartimento di Fisica e Astronomia, Universit\`a di Catania,\\
and CNISM, UdR Catania, and INFN, Sez. Catania,\\
64, Via S. Sofia, I-95129 Catania, Italy.
\end{em}
\end{center}

\medskip

\begin{abstract}
In the centennial of Ettore Majorana's birth (1906---1938?), we re-examine some
aspects of his fundamental scientific production in atomic and molecular
physics, including a not well known short communication. There, Majorana
critically discusses Fermi's solution of the celebrated Thomas-Fermi equation
for electron screening in atoms and positive ions. We argue that some of
Majorana's seminal contributions in molecular physics already prelude to the
idea of exchange interactions (or Heisenberg-Majorana forces) in his later works
on theoretical nuclear physics. In all his papers, he tended to emphasize the
symmetries at the basis of a physical problem, as well as the limitations,
rather than the advantages, of the approximations of the method employed.

\medskip

\end{abstract}

\section{Introduction}

Ettore Majorana's most famous, seminal contributions are certainly those on the
relativistic theory of a particle with an arbitrary instrinsic angular momentum
\cite{Majorana:32}, on nuclear theory \cite{Majorana:33}, and on the symmetric
theory of the electron and the positron \cite{Majorana:37}. In particular, the
latter paper already contains the idea of the so-called Majorana neutrino
\cite{Majorana:37}, as has been correctly emphasized \cite{Amaldi:66}. The quest
for Majorana neutrinos is still the object of current fundamental research (see,
\emph{e.g.,} Ref.~\cite{Sapienza:05}, and Ref.~\cite{Bettini:04} for a general
overview).

In this note, we would like to reconsider two more papers by Majorana
\cite{Majorana:29,Majorana:31}, both on atomic and molecular physics, and show
how they are precursor to his theoretical work on the exchange nuclear forces,
the so-called Heisenberg-Majorana forces \cite{Majorana:33}. We will also try
and emphasize his critical sense and great ability to catch the relevant
physical aspects of a given problem, beyond his celebrated mathematical skills,
as witnessed by contemporaries and colleagues who met him personally
\cite{Amaldi:68,Majorana:87,Bonolis:02} (see especially Ref.~\cite{Recami:99}
for more references).

Both Amaldi and Segr\`e have provided us with a vivid account of Majorana's
first meeting with Enrico Fermi. Majorana and Fermi first met in 1928 at the
Physical Institute in Via Panisperna, Rome. At that time, Fermi was working on
his statistical model of the atom, known nowadays as the Thomas-Fermi model,
after the names of the two authors who derived it independently
\cite{Thomas:26,Fermi:27,Fermi:28}. Such a model provides an approximate
alternative to solving Schr\"odinger equation \cite{March:75a}, and paved the
way to density functional theory \cite{Pucci:86a}.

\section{Thomas-Fermi model}

Within Thomas-Fermi approximation, the electronic cloud surrounding an atom is
described in terms of a completely degenerate Fermi gas. Following
Ref.~\cite{March:75a}, one arrives at a \emph{local} relation between the
electron density $\rho(\br)$ at position $\br$ with respect to the nucleus, and
the momentum $\pF (\br)$ of the fastest electron (Fermi momentum), as
\begin{equation}
\rho(\br) = 2 \cdot \frac{4\pi}{3} \pF^3 (\br),
\label{eq:Fermi}
\end{equation}
where the factor of two takes into account for Pauli exclusion. In
Eq.~(\ref{eq:Fermi}), the Fermi momentum $\pF (\br)$ depends on position $\br$
through the self-consistent potential $V(\br)$ as
\begin{equation}
\pF^2 (\br) = 2 m [\EF - V(\br)] ,
\label{eq:Vr}
\end{equation}
where $\EF$ is the Fermi energy, and $m$ is the electron mass.
Fermi energy $\EF$ is then determined via the normalization condition
\begin{equation}
\int \rho(\br) \, d^3 \br = N ,
\end{equation}
where $N$ is the total electron number, equalling the atomic number $Z$ for a
neutral atom.
Inserting Eq.~(\ref{eq:Vr}) into Eq.~(\ref{eq:Fermi}), making use of Poisson
equation, and introducing Thomas-Fermi screening factor $\phi$ through
\begin{equation}
V(\br) - \EF = - \frac{Ze^2}{r} \phi(\br),
\label{eq:Vsc}
\end{equation}
one derives the adimensional Thomas-Fermi equation for a spherically symmetric
electron distribution,
\begin{equation}
\frac{d^2 \phi}{dx^2} = \frac{\phi^{3/2}}{x^{1/2}} ,
\label{eq:ThF}
\end{equation}
where
\begin{equation}
r = b x ,
\end{equation}
and $b$ sets the length scale as
\begin{equation}
b = \frac{1}{4} \left( \frac{9\pi^2}{2Z} \right)^{1/3} a_0 =
\frac{0.8853}{Z^{1/3}} a_0 ,
\label{eq:b}
\end{equation}
with $a_0$ the Bohr radius.

Eq.~(\ref{eq:ThF}) is `universal', in the sense that the sole dependence on the
atomic number $Z$ comes through Eq.~(\ref{eq:b}) for $b$. Once
Eq.~(\ref{eq:ThF}) is solved, the self-consistent potential for the particular
atom under consideration is simply obtained by scaling all distances with $b$.

\section{Asymptotic behaviour of the solution to Thomas-Fermi equation}

Fermi endeavoured to solve Eq.~(\ref{eq:ThF}) analytically without success. On
the occasion of his first meeting with Majorana, Enrico succintly exposed his
model to Ettore, and Majorana got a glimpse of the numerical results he had
obtained over a week time, with the help of a primitive calculator. The day after
Majorana reappeared and handled a short note to Fermi, where he had jotted down
\emph{his} results. Majorana was amazed that Fermi's results coincided with his
own.

How could Majorana solve Eq.~(\ref{eq:ThF}) numerically in such a short time
without the help of any calculator? Various hypotheses have been proposed. Did
he find an analytical solution? At any rate, there are no physically acceptable
analytical solutions to Eq.~(\ref{eq:ThF}) in the whole range $0\leq x <
+\infty$. The only analytical solution,
\begin{equation}
\phi(x) = \frac{144}{x^3} ,
\label{eq:Sommerfeld}
\end{equation}
would have been found later by Sommerfeld in 1932 \cite{Sommerfeld:32}, and is
physically meaningful only asymptotically, for $x\gg 1$.

The most likely hypothesis is probably that of Esposito \cite{Esposito:02}, who,
together with other authors \cite{Esposito:03}, has found an extremely original
solution to Eq.~(\ref{eq:ThF}) in Majorana's own notes (see also
Ref.~\cite{DiGrezia:04}). The method devised by Majorana leads to a
semi-analytical series expansion, obeying both boundary conditions for a neutral
atom
\begin{subequations}
\label{eq:Fbound}
\begin{eqnarray}
\phi(0) &=& 1 , \\
\phi(\infty) &=& 0 .
\label{eq:Fbound:b}
\end{eqnarray}
\end{subequations}

In a recent work, Guerra and Robotti \cite{Guerra:05} have rediscovered a
not well known short communication by Majorana, entitled \emph{Ricerca di
un'espressione generale delle correzioni di Rydberg, valevole per atomi neutri o
ionizzati positivamente} (\emph{Quest for a general expression of Rydberg
corrections, valid for either neutral or positively ionized atoms})
\cite{Majorana:29}. In that work, perhaps in the attempt of improving the
asymptotic behaviour of the solution to Thomas-Fermi equation, Ettore requires
that the potential vanishes for a certain finite value of $x$, say $x_0$,
\emph{both} for neutral atoms and for positive ions. He writes the
self-consistent potential as
\begin{equation}
V (r ) = \frac{Ze}{r} \phi + C,
\label{eq:VscM}
\end{equation}
where, for an atom positively ionized $n$ times ($n=Z-N$), the constant $C$
equals
\begin{equation}
C= \frac{n+1}{bx_0} e ,
\end{equation}
where
\begin{equation}
b = 0.47 \, \frac{1}{Z^{1/3}} \left( \frac{Z-n}{Z-n-1} \right)^{2/3}
\,\mathrm{\mbox{\AA{}}} ,
\label{eq:bM}
\end{equation}
and the boundary conditions to Eq.~(\ref{eq:ThF}) now read
\begin{subequations}
\label{eq:Mbound}
\begin{eqnarray}
\phi(0) &=& 1, \\
\phi(x_0 ) &=& 0, \\
-x_0 \phi^\prime (x_0 ) &=& \frac{n+1}{Z} .
\end{eqnarray}
\end{subequations}

One immediately notices that, due to the new boundary conditions,
Eq.~(\ref{eq:VscM}) does not reduce to Eq.~(\ref{eq:Vsc}) even for $n=0$,
\emph{i.e.} for a neutral atom. In other words, Majorana does not consider the
potential $V(\br)$ in a generic location of the electron cloud, but the
effective potential acting on a single electron, thus excluding the interaction
of an electron with itself.

Probably, owing to his profound critical sense (let us remind that his
colleagues in the Panisperna group nicknamed him the `Great Inquisitor'),
Majorana must have not excessively relied on his own solution
\cite{Esposito:02}, which however reproduced the numerical solution of
Thomas-Fermi equation quite accurately. Probably, Majorana was looking for a
solution which should not decrease so slowly as $x\to\infty$, as
Eq.~(\ref{eq:Sommerfeld}) does.

\begin{figure}[t]
\centering
\includegraphics[height=0.8\columnwidth,angle=-90]{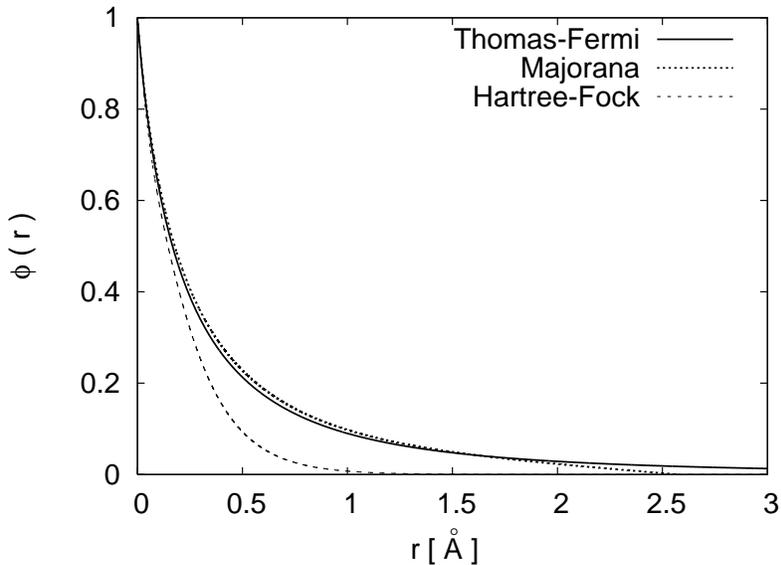}
\caption{Thomas-Fermi screening factor $\phi$ for the self-consistent potential
of a neutral Ne atom ($Z=N=10$). Solid line is Fermi's solution, dashed line is
Majorana's solution, while the light dashed line has been obtained within
Hartree-Fock approximation.}
\label{fig:neutral}
\end{figure}

In Fig.~\ref{fig:neutral} we report Thomas-Fermi screening factor $\phi$ as a
function of $r$ for a neutral Ne atom ($Z=N=10$). The solid line refers to
Fermi's numerical solution, with boundary conditions given by
Eq(s).~(\ref{eq:Fbound}), the dashed line refers to Majorana's solution, with
boundary conditions given by Eq(s).~(\ref{eq:Mbound}) with $n=0$, while the light
dashed line has been obtained within the Hartree-Fock approximation (see
App.~\ref{app:HF} for a derivation). As it can be seen, Majorana's solution
introduces only a minor correction to Fermi's solution at finite $x$ values, but
is strictly zero for $x\geq x_0$.

In his work on positive ions \cite{Fermi:30}, Fermi considers a potential
vanishing at a finite value $x=x_0$. However, instead of Eq(s).~(\ref{eq:Mbound}),
he employs the boundary conditions
\begin{subequations}
\begin{eqnarray}
\phi(0) &=& 1,\\
-x_0 \phi^\prime (x_0 ) &=& \frac{n}{Z} ,
\label{eq:Fbound:c}
\end{eqnarray}
\end{subequations}
which in particular imply Eq.~(\ref{eq:Fbound:b}) in the case $n=0$,
corresponding to a neutral atom.

\begin{figure}[t]
\centering
\includegraphics[height=0.8\columnwidth,angle=-90]{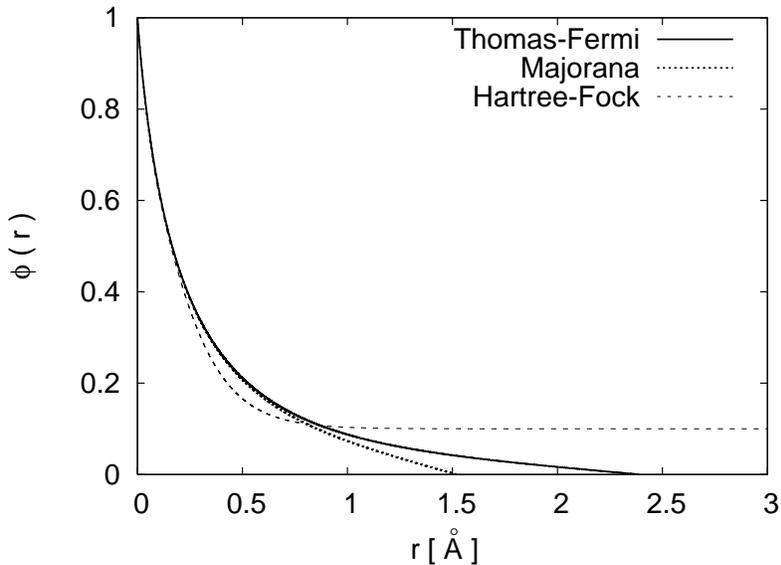}
\caption{Thomas-Fermi screening factor $\phi$ for the self-consistent potential
of the Ne$^+$ ion ($Z=10$, $N=9$). Solid line is Fermi's solution, dashed line is
Majorana's solution, while the light dashed line has been obtained within
Hartree-Fock approximation.}
\label{fig:ion}
\end{figure}

In Fig.~\ref{fig:ion}, we again report Thomas-Fermi screening factor $\phi$ as a
function of $r$ according to Fermi, Majorana, and Hartree-Fock, respectively,
but now for a positively ionized Ne atom, Ne$^+$ ($Z=10$, $N=9$, $n=1$).
Majorana's solution again differs but marginally from Fermi's solution, but
while for a neutral Ne atom Fermi's solution decreases too slowly, it decreases
too rapidly for Ne$^+$.

Here, we are not disputing whether Majorana's note, Ref.~\cite{Majorana:29},
should be considered as a `full' paper \cite{Esposito:05}, nor do we want to
undervalue the importance of the contribution analyzed in
Ref.~\cite{Esposito:02}. We would rather like to emphasize that Majorana was
conscious that his correction\footnote{Fl\"ugge \cite{Fluegge:74} erroneously
attributes this correction to Amaldi. Probably, he was only aware of Fermi and
Amaldi's final work, Ref.~\cite{Fermi:34}.} did not lead to substantial
modifications to Fermi's solution of Eq.~(\ref{eq:ThF}), including in the
asymptotic limit ($x\gg 1$) \cite{Pucci:82}.

Ettore never published anything else on this subject.

\section{Helium molecular ion}

In his successive work \cite{Majorana:31}, Majorana deals with the formation of
the molecular ion He$_2^+$. There again, Majorana demonstrates his exceptional
ability to focus on the main physical aspects of the problem, while showing the
limitations of his own theoretical approximations. He immediately observes that
the problem is more similar to the formation of the molecular ion H$_2^+$, than
to the reaction $\mathrm{He} + \mathrm{H}$. The most relevant forces, especially
close to the equilibrium distance, are therefore the \emph{resonance} forces,
rather than the polarization ones. By \emph{exchanging} the two nuclei, the
system remains unchanged. Majorana makes then use of the method of Heitler and
London \cite{Heitler:27}, and emphasizes the importance of inversion symmetry
with respect to the middle point between the nuclei, set at a distance $R$
apart.

Heitler and London \cite{Heitler:27} introduced a relatively simple expression
for the wave-function $\Psi$ of the two electrons in a hydrogen molecule H$_2$ in
terms of the wave-functions $\varphi$ and $\psi$ of one electron in the atomic orbital
corresponding to atom $a$ and $b$, respectively:
\begin{equation}
\Psi(1,2) = \varphi(1) \psi(2) \pm \psi(1) \varphi(2),
\label{eq:HL}
\end{equation}
where $1$ and $2$ denote the coordinates of the two electrons, respectively.
The wave-function $\Psi_S$, corresponding to the choice of the plus sign in
Eq.~(\ref{eq:HL}), is symmetric with respect to the exchange of the coordinates
of both electrons and nuclei, while $\Psi_A$ (minus sign in Eq.~(\ref{eq:HL}))
is antisymmetric. The full wave-function is globally anti-symmetric, but here we
are neglecting its spin part, since the Hamiltonian is spin independent.

Despite its simplicity, the success of Heitler-London approximation relies on
the fact that it explained the stability of the H$_2$ molecule, and could
reproduce with remarkable accuracy the dependence of the total electronic energy
$E_I$ on the internuclear distance $R$. One obtains the attractive solution in
correspondence with the eigenfunction $\Psi_S$. It is relevant to stress at this
point that, if one had considered only $\varphi(1) \psi(2)$, or $\psi(1) \varphi(2)$, in
Eq.~(\ref{eq:HL}), the agreement with experimental data would have been rather
poor. Therefore, the resonance or exchange term is quite decisive for
establishing the chemical bond.

Heitler-London theory is even more accurate than the method of molecular
orbitals \cite{Hund:28,Mulliken:28,Mulliken:28a,Hueckel:30} (see, \emph{e.g.,}
Ref.~\cite{Coulson:52} for a more detailed discussion), which in addition to
Eq.~(\ref{eq:HL}) takes into account also for the ionic-like configurations
\begin{equation}
\varphi(1)\varphi(2) \quad \mbox{and} \quad \psi(1)\psi(2),
\label{eq:molorb}
\end{equation}
corresponding to having both electrons on atom $a$, or $b$, respectively,
on the same footing and with equal weights as the terms in Eq.~(\ref{eq:HL}).
However, the theory can be improved by adding to Eq.~(\ref{eq:HL}) the two
contributions in Eq.~(\ref{eq:molorb}) with appropriate weights, to be
determined variationally. 

As in Heitler-London, in order to study the case of He$_2^+$, also Majorana
starts from the asymptotic solution, namely for large values of $R$.
In the case of H$_2$, for large values of $R$, it is very unlikely that both
electrons reside on the same nucleus.
Similarly, in the case of He$_2^+$, Majorana neglects the possibility that all
three electrons be located on the same nucleus.
Ettore then proceeds by writing the unperturbed eigenfunctions for the three
electrons (labeled 1, 2, 3 below) in He$_2^+$ as
\begin{subequations}
\begin{alignat}{3}
A_1 &= \varphi_1 \Psi_{23}, & \qquad B_1 &= \psi_1 \Phi_{23}, \\
A_2 &= \varphi_2 \Psi_{31}, & \qquad B_2 &= \psi_2 \Phi_{31}, \\
A_3 &= \varphi_3 \Psi_{12}, & \qquad B_3 &= \psi_3 \Phi_{12}, 
\end{alignat}
\end{subequations}
where $\Phi$ and $\varphi$ denote the wave-functions of the neutral and ionized
$a$ atom, respectively, while $\Psi$ and $\psi$ denote the analogous
wave-functions for atom $b$.
Evidently, $A_2$ and $A_3$ can be obtained from $A_1$ by permuting the
electrons, and the $B$'s from the $A$'s by \emph{exchanging} the nuclei.

The interaction between the atoms mixes all these wave-functions, but by means
of general symmetry considerations, first introduced by Hund \cite{Hund:27}, as
well as of inversion symmetry and of Pauli exclusion principle, Majorana
concludes that the only acceptable wave-functions are
\begin{subequations}
\begin{eqnarray}
y_1 &=& A_1 - A_2 + B_1 - B_2 ,\\
y_2 &=& A_1 - A_2 - B_1 + B_2 ,
\label{eq:y2}
\end{eqnarray}
\end{subequations}
which are antisymmetric in 1 and 2. In particular, $y_2$ corresponds to the
$(1s\sigma)^2\, 2 p\sigma\, ({}^2 \Sigma)$ configuration, \emph{viz.} the
bonding solution for the He$_2^+$ molecular ion. The latter configuration is
characterized by two electrons in the $\sigma$ molecular orbital built from the
two $1s$ atomic orbitals, one electron in the $\sigma$ molecular orbital built
from the $2p$ atomic orbitals, as well as by a value of the total orbital
angular momentum $L=0$, and by a value of the total spin $S=+\frac{1}{2}$. The
wave-function Eq.~(\ref{eq:y2}) clearly shows that the ground state is a
resonance between the configurations $\mathrm{He\colon - He\cdot}$ and
$\mathrm{He\cdot - He\colon}$, where each dot denotes the presence of one
electron on the $a$ or $b$ atom.

In order to perform the calculation of the interaction terms explicitly, making
use of analytic expressions, one can take the ground state of the helium atom as
the product of two hydrogenoid wave-functions. However, it is well known that
the result is greatly improved if, instead of taking the bare charge $Z=2$ of
the He nucleus, an effective nuclear charge $Z_{\mathrm{eff}}$ is introduced, to
be determined variationally. The fundamental effect here taken into account by
Majorana is that of screening: In an atom with many electrons, each electron
sees the nuclear charge $Ze$ as slightly \emph{attenuated} by the presence of
the remaining electrons.

The concept of an effective nuclear charge, already introduced for the helium
atom, had been extended by Wang \cite{Wang:28} to the hydrogen molecule.
Probably Majorana was not aware of Wang's work, since he does not refer to it in
his 1931 paper. In any case, Majorana is the first one to make use of such a
method for He$_2^+$. In making reference to his own work \cite{Pauling:33},
where $Z_{\mathrm{eff}}$ is used as a variational parameter for He$_2^+$,
Pauling reports in a footnote\footnote{See footnote on p.~359 of
Ref.~\cite{Pauling:35a}.}

\begin{quote}
``The same calculation with $Z_{\mathrm{eff}}$ given the fixed value 1.8 was
made by E. Majorana \cite{Majorana:31}.'' 
\end{quote}

\noindent
The variational value obtained by Pauling for $Z_{\mathrm{eff}}$ is 1.833.

By making use of his results, Majorana evaluates the equilibrium internuclear
distance as $d=1.16$~\AA, in good agreement with the experimental value
1.087~\AA. He can then estimate the vibrational frequency as $n=1610$~cm$^{-1}$,
which he compares with the experimental value $1628$~cm$^{-1}$. Majorana
concludes his paper by stating \cite{Majorana:31} that his own result is 

\begin{quote}
``\emph{casually} in perfect agreement with the experimentally determined
value'' 
\end{quote}

\noindent
(our italics). Any other author would have emphasized such a striking agreement
as a success of his own method, whereas Majorana rather underlines the drawbacks
of his own approximations.

\begin{figure}[t]
\centering
\includegraphics[height=0.8\columnwidth,angle=-90]{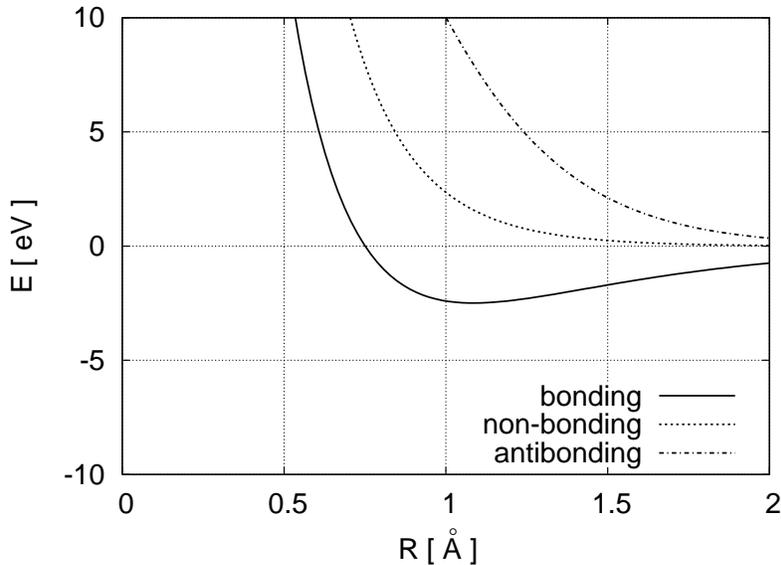}
\caption{Variational energies of the molecular ion He$_2^+$, as a function of
the internuclear distance $R$. Solid line refers to the symmetric wave-function
in Eq.~(\ref{eq:HL}), dashed-dotted line to the antisymmetric one, while dashed
line refers to the `non-bonding' case, where position exchange is neglected.
Redrawn after Ref.~\cite{Pauling:33} (see also Ref.~\cite{Pauling:35}).}
\label{fig:Pauling}
\end{figure}

We would like to remind that he also estimates the minimum energy, \emph{i.e.}
the dissociation energy, finding the value $E_{\mathrm{min}} = -1.41$~eV, but he
had no available experimental data to compare with, at that time. However, he is
not satisfied with such a result and collects \cite{Majorana:31}

\begin{quote}
``all the errors of the method under the words `\emph{polarization} forces','' 
\end{quote}

\noindent
which he estimates for very distant nuclei using the polarizability of the
neutral He atom. He then finds $E_{\mathrm{min}} = -2.4$~eV. More recent
theoretical calculations, using the method of configuration interactions
\cite{Ackermann:91} or ab~initio variational methods \cite{Reagan:63}, have
estimated the value $E_{\mathrm{min}} = -2.47$~eV. The experimental value has
been accurately determined quite recently \cite{Coman:99} as $E_{\mathrm{min}} =
-2.4457 \pm 0.0002$~eV. We are not claiming that Ettore's result is more
accurate than the theoretical results mentioned above. However, he certainly
understood the essential physical effects for that system, and made use of
appropriate approximations to estimate them. In particular, it is interesting
how he emphasizes the quest for the \emph{symmetries} of the system (see the
translation of a paper by Majorana in Ref.~\cite{Esposito:03}). As in the case
of H$_2$, also for He$_2^+$ it is essential to include the position exchange
term between He and He$^+$, in order to have chemical bonding, as it can be seen
in Fig.~\ref{fig:Pauling}, redrawn after Ref.~\cite{Pauling:35}. If one had
neglected the resonance $\mathrm{He} \colon \mathrm{He}^+ \rightleftharpoons
\mathrm{He}^+ \colon \mathrm{He}$ (see dashed line in Fig.~\ref{fig:Pauling}),
chemical bonding would have been impossible.

\section{The discovery of the neutron}

Rutherford's pioneering work \cite{Rutherford:63} paved the way not only to
Bohr's atomic model, but also to nuclear physics.

In 1930 Bothe and Becker \cite{Bothe:30}, like Rutherford, employed
$\alpha$~particles against a berillium target in a scattering experiment. They
observed the emission of a very penetrating radiation, which they interpreted as
$\gamma$~rays. In successive experiments, Ir\`ene Curie and Frederic Joliot
\cite{Curie:32,Curie:32a}, her husband, developed further these experiments, but
they arrived at similar conclusions. According to Emilio Segr\`e's account
\cite{Segre:70}, Majorana thus commented the Joliots' results: 

\begin{quote}
``They haven't realized they have discovered the neutral proton.''
\end{quote}

At this point we should remind that at that time it was believed that the
nucleus was composed by protons and electrons. It was Chadwick
\cite{Chadwick:32} who soon after demonstrated that the radiation emitted in the
Joliots' experiments was made up by neutral particles, whose mass is very close
to the proton's mass. It was probably Fermi \cite{Segre:70} who first
distinguished between the \emph{neutrinos} conjectured by Pauli, and the
\emph{neutrons} discovered by Chadwick.

Meanwhile, Majorana developed a theory of the nucleus containing protons and
neutrons and then, according to Segr\`e \cite{Segre:70}, 

\begin{quote}
``he analyzed, as far as it was possible, the nuclear forces on the basis of the
available experimental results, and he estimated the binding energies of the
lightest nuclei. When he presented his work to Fermi and ourselves, we
immediately recognized its importance. Fermi encouraged Majorana to publish his
own results, but Majorana refused to do so, saying they were yet too
incomplete.''
\end{quote}

\noindent 
More than that, when Fermi asked Majorana whether he could make reference to his
results during a forthcoming conference in Paris, Ettore mockingly replied he
would agree, provided the reference was attributed to an old professor of
electrochemistry, who was also going to attend the same conference.
Obviously, Fermi could not accept Majorana's condition, and no reference was
then made to his results during the conference.

Meanwhile, people started feeling the lack of a theory of nuclear forces,
conveniently taking into account for the presence of both protons and nucleons
in the nucleus. But where to begin with?

\section{Heisenberg-Majorana forces}

\begin{figure}[t]
\centering
\includegraphics[width=0.9\columnwidth]{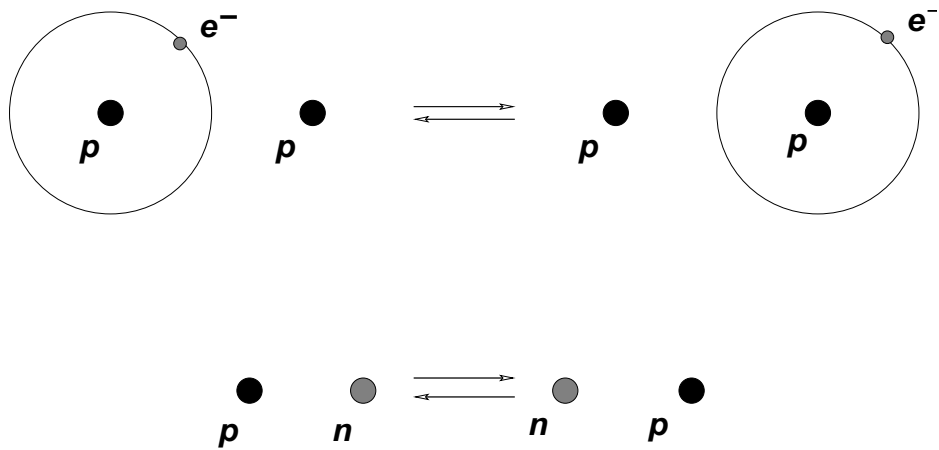}
\caption{Exchange interactions. Resonant forms in the hydrogen molecular ion,
$\mathrm{H} \colon \mathrm{H}^+ \protect\rightleftharpoons \mathrm{H}^+ \colon
\mathrm{H}$ (upper row), and in the proton-nucleon pair inside a nucleus,
$p\colon n \protect\rightleftharpoons n \colon p$ (lower row).}
\label{fig:exchange}
\end{figure}

To this aim, in three fundamental contributions
\cite{Heisenberg:32,Heisenberg:33,Heisenberg:33a}, Heisenberg assumed hydrogen
molecular ion H$_2^+$ as a model. He recognizes that the most important nuclear
forces are not the polarization forces among the neutrons, or Coulombic
repulsion among protons, but the exchange forces between protons and neutrons.

Heisenberg emphasizes that neutrons obey to Fermi statistics. Moreover, since a
neutron possesses spin $\frac{1}{2}\hbar$, it cannot be simply thought of as
composed of a proton plus an electron, unless the latter has zero spin, when
inside a neutron.\footnote{Besides considerations concerning the spin, such a
model would require an enormous amount of energy to localize the electron within
the neutron \cite{Heisenberg:33}.} A neutron is an elementary particle \emph{per
se.} The  interactions postulated by Heisenberg are characterized by the
exchange of both position coordinates and spins of the two nucleons.

\begin{figure}[t]
\centering
\includegraphics[height=0.9\columnwidth,angle=-90]{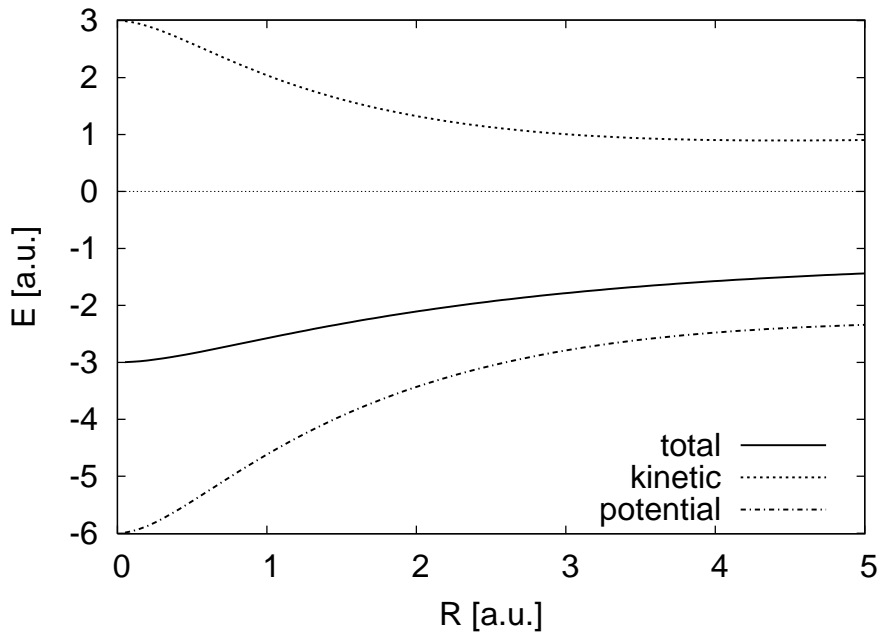}
\caption{Kinetic, potential, and total energies for the ground state of H$_2^+$,
excluding nuclear repulsion, within the linear combination of atomic orbitals
(LCAO) approximation. Cf. Fig.~2.4 in Ref.~\cite{Slater:63}, where the same
quantities have been obtained within a variational method.}
\label{fig:Jr}
\end{figure}

Similarly, Majorana assumed that the fundamental nuclear forces are of exchange
nature between protons and neutrons. However, he fully exploits the analogy with
H$_2^+$ (see Fig.~\ref{fig:exchange}), regardless of spin.\footnote{Current
literature usually employs the formalism of isotopic spin to describe the
exchange character of the nuclear forces. However, as noted by Blatt and
Weisskopf \cite{Blatt:52}, this is equivalent to a description which makes use
of the forces of Bartlett, Heisenberg, Majorana, and Wigner.}

Let $\br_1$, $\sigma_1$ and $\br_2$, $\sigma_2$ stand for the position and spin
coordinates of the first and the second nucleon, respectively, and let
$\psi(\br_1 , \sigma_1 ; \br_2 , \sigma_2 )$ be the wave-function for a given
nucleon pair \cite{Blatt:52}. Then Heisenberg exchange $P^H$ implies
\begin{equation}
P^H \psi(\br_1 , \sigma_1 ; \br_2 , \sigma_2 ) =
\psi(\br_2 , \sigma_2 ; \br_1 , \sigma_1 ),
\end{equation}
whereas Majorana exchange $P^M$ implies
\begin{equation}
P^M \psi(\br_1 , \sigma_1 ; \br_2 , \sigma_2 ) =
\psi(\br_2 , \sigma_1 ; \br_1 , \sigma_2 ).
\end{equation}
In Majorana's own notation (apart from a minus sign here included in the
definition of $J(r)$), the exchange interaction then reads
\cite{Majorana:33}
\begin{equation}
(Q^\prime , q^\prime | J | Q^{\prime\prime} , q^{\prime\prime} ) = J(r) \delta
(q^\prime - Q^{\prime\prime} ) \delta (q^{\prime\prime} - Q^\prime ),
\label{eq:Jr}
\end{equation}
where $Q$ and $q$ are the position coordinates of the neutron and the proton,
respectively, and $r=|q^\prime - Q^\prime |$ is their relative distance.
Majorana then plots a qualitative sketch of $J(r)$ (cf. Fig.~2 in
Ref.~\cite{Majorana:33}), which closely resembles the behaviour of the potential
energy in H$_2^+$, when the internuclear repulsion is neglected
(Fig.~\ref{fig:Jr}).

In the same paper \cite{Majorana:33}, in addition to his knowledge of molecular
physics, Majorana fully exploits also his acquaintance with the atomic
statistical model. Indeed, he defines the nuclear density as
\begin{equation}
\rho = \frac{8\pi}{3h^3} (P_n^3 + P_p^3 ),
\end{equation}
in complete analogy with Eq.~(\ref{eq:Fermi}), where $P_n$ and $P_p$ are the 
Fermi momenta of neutrons and protons, respectively. From this model, he
derives an asymptotic expression ($\rho\to\infty$) for the exchange energy per
particle,
\begin{equation}
\left.a(\rho)\right|_{\rho\to\infty} = - \frac{n_2}{n_1 + n_2} J(0) ,
\end{equation}
where $n_1$ and $n_2$ are the numbers of neutrons and protons, respectively.
As in Thomas-Fermi model, the kinetic energy per particle, $t$ say, is given by
\begin{equation}
t\propto \rho^{2/3} .
\end{equation}
From the competition between kinetic and potential energy, the total energy
attains a minimum as a function of $r$ (cf. Fig.~1 in Ref.~\cite{Majorana:33}).

Majorana's model explains two fundamental properties of nuclear physics
\cite{Blatt:52}: (a) the density of nucleons is about the same for all nuclei
(\emph{density saturation}); (b) the binding energy per nucleon is about the
same for all nuclei (\emph{binding energy saturation}).

\section{Concluding remarks}

In some of his fundamental papers, Majorana mainly focussed on the asymptotic
properties of the potential and of the wave-function of an atomic or molecular
system. This is clearly demonstrated in his work on helium molecular ion,
He$_2^+$ \cite{Majorana:31}. On the basis of his hypercritical spirit, Majorana
was probably unsatisfied with the asymptotic behaviour of the screening factor
$\phi$ within Thomas-Fermi model, but his note \cite{Majorana:29} is too short
to confirm that. What we can certainly emphasize is his taste for the quest of
symmetries, and their relevance to determine the main properties of a physical
system \cite{Esposito:05}. This led him to demonstrate that the exchange
symmetry is essential to the formation of the chemical bond. Exchange symmetry
is also central in his model of the nuclear forces.

The quest for symmetries is evident in his famous work on the symmetrical theory
of the electron and the positron \cite{Majorana:37}. There, he notes that

\begin{quote}
``all devices suggested to endow the theory \cite{Dirac:24} with a symmetric
formulation, without violating its contents, are not completely satisfactory.
[\ldots] It can be demonstrated that a quantum theory of the electron and the
positron can be formally symmetrized completely by means of a new quantization
procedure. [\ldots] Such a procedure not only endows the theory with a fully
symmetric formulation, but also allows one to construct a substantially new theory
for chargeless [elementary] particles (neutrons and hypothetical neutrinos).''
\end{quote}

Several important experiments \cite{Sapienza:05,Bettini:04} are currently under
way to observe the `Majorana neutrino'.

\subsection*{Acknowledgements}

The authors are grateful to Professor M. Baldo for useful comments and for
carefully reading the manuscript before publication, and to Professor N.~H.
March for close collaboration and many stimulating discussions over the general
area embraced by this note. The authors also acknowledge helpful discussions
with Dr. G. Piccitto.

\appendix

\section{Thomas-Fermi screening factor\\ within Hartree-Fock approximation}
\label{app:HF}

In order to critically assess the accuracy of Fermi's and Majorana's approximate
solutions for the atomic screening factor $\phi$ in Thomas-Fermi model, let us
briefly derive it within the Hartree-Fock self-consistent approximation. The
solution of Hartree-Fock equations enables one to determine the (spherically
symmetric) radial electron density
\begin{equation}
D(r) = 4\pi r^2 \rho(r),
\label{eq:distr}
\end{equation}
normalized to the total electron number as
\begin{equation}
\int_0^\infty D(r)\, dr = N
\label{eq:norm}
\end{equation}
(see, \emph{e.g.,} Ref.~\cite{Bransden:03}). 
The radial electron density $D(r)$ of a neutral Ne atom is characterized by two
peaks, referring to the $1s$ and $2s\,2p$ shells, respectively (cf. Fig.~8.6 in
Ref.~\cite{Bransden:03}). In the case of Ne$^+$, such peaks are slightly shifted
at smaller values of $r$. Although $D(r)$ is always strictly different from zero
over the whole $r$ range, it is an exponentially decreasing function of $r$,
with $D(r)\approx 0$ roughly defining the atomic (respectively, ionic) radius.

By relating the electric field $|\vec{E} | = (1/e) \partial V / \partial r$
corresponding to the self-consistent potential, Eq.~(\ref{eq:Vsc}), to that
generated by the nucleus and the electron cloud within a distance $r$ from the
nucleus, by Gauss law,
one finds
\begin{subequations}
\begin{eqnarray}
\label{eq:phip}
\phi^\prime &=& \frac{1}{r} \left[ \phi - 1 + \frac{1}{Z} \int_0^r D(r^\prime ) \,
dr^\prime \right] ,\\
\phi(0) &=& 1 ,
\end{eqnarray}
\end{subequations}
where a prime here refers to derivation with respect to $r$.

Within Thomas-Fermi approximation, $\phi(r_0 )=0$, where $r_0$ is the ionic
radius, and the integration in the normalization condition, Eq.~(\ref{eq:norm}),
should actually be performed up to $r=r_0$. Then, Eq.~(\ref{eq:phip}) reduces to
Fermi's boundary condition, Eq.~(\ref{eq:Fbound:c}).

Within Hartree-Fock approximation, $D(r)$ is in general nonzero at any finite
$r$. However, as $r\to\infty$, the potential experienced by a test charge is
that of a charge $(Z-N)e$, \emph{i.e.} $V(r) - \EF \sim - (Z-N)e^2/ r$.
Comparing such an asymptotic behaviour of the potential with the definition of
the screening function $\phi$ in Eq.~(\ref{eq:Vsc}), one has
\begin{equation}
\lim_{r\to\infty} \phi(r) = 1- \frac{N}{Z}.
\end{equation}
On the other hand, making use of the latter result, from Eq.~(\ref{eq:phip}) it
follows that
\begin{equation}
\lim_{r\to\infty} r\phi^\prime (r) = 0,
\label{eq:phipinfty}
\end{equation}
thus implying that $\phi^\prime (r)$ vanishes as $r\to\infty$ more rapidly that
$1/r$ (in fact, it vanishes exponentially).

Finally, from Poisson equation, $-\nabla^2 V = 4\pi e^2 [-Z\delta(r) + \rho(r)
]$, for a given electron charge distribution, Eq.~(\ref{eq:distr}), one obtains
\begin{equation}
\phi^{\prime\prime} = \frac{1}{Z} \frac{D(r)}{r}
\end{equation}
at any $r>0$.
Integrating once between $r$ and $\infty$, and making use of the limiting value
Eq.~(\ref{eq:phipinfty}), one obtains
\begin{equation}
\phi^\prime (r) = - \frac{1}{Z} \int_r^\infty \frac{D(r^\prime )}{r^\prime} \,
dr^\prime ,
\end{equation}
which combined with Eq.~(\ref{eq:phip}) yields the desired screening factor
\begin{equation}
\phi(r) = 1 - \frac{1}{Z} \int_0^r D(r^\prime ) \, dr^\prime - \frac{r}{Z}
\int_r^\infty \frac{D(r^\prime )}{r^\prime} \, dr^\prime
\label{eq:phiHF}
\end{equation}
in terms of the Hartree-Fock self-consistent radial density $D(r)$.
In particular, Eq.~(\ref{eq:phiHF}) manifestly fulfills the boundary conditions
\begin{subequations}
\begin{eqnarray}
\phi(0) &=& 1\\
\phi(\infty) &=& 1 - \frac{N}{Z} .
\end{eqnarray}
\end{subequations}
In Fig(s).~\ref{fig:neutral} and \ref{fig:ion}, light dashed lines represent
Eq.~(\ref{eq:phiHF}), with $D(r)$ numerically obtained within Hartree-Fock
self-consistent approximation for Ne and Ne$^+$, respectively.

\begin{small}
\bibliographystyle{fpl}
\bibliography{a,b,c,d,e,f,g,h,i,j,k,l,m,n,o,p,q,r,s,t,u,v,w,x,y,z,zzproceedings,Angilella,notes}
\end{small}

\end{document}